\documentclass[twocolumn,twoside,slac_two]{revtex4}
\usepackage{graphicx}
\usepackage{fancyhdr}
\usepackage{hyperref}
\hypersetup{
    colorlinks=true,
    urlcolor=magenta,
}
\begin{document}

\title{Computation of complex ion production due to cosmic rays during the Halloween sequence of GLEs on October-November 2003}

%

\author{A. Mishev}
\affiliation{Space Climate Research Unit, University of Oulu, Oulu, Finland}
\author{P.I.Y. Velinov}
\affiliation{Institute for Space Research and Technology, Bulgarian Academy of Sciences, Sofia, Bulgaria}

\begin{abstract}
The possible effect of solar variability, accordingly cosmic rays variation on atmospheric physics and chemistry is highly debated over the last years. According to several recent models  the induced by cosmic rays atmospheric ionization plays a key role in several different processes. At recent, an apparent effect on minor constituents and aerosols over polar regions during major solar proton events was observed. The ion production rate during  ground level enhancements is a superposition of the contribution of cosmic rays with galactic and solar origin. The solar cycle 23 provided several strong ground level enhancements. The period of end October -  beginning of November 2003 was characterized by a strong cosmic ray variability, namely a sequence of three GLEs was observed. In addition, there were several Forbush decreases, which led to a suppression of galactic cosmic ray flux. As a consequence the cosmic ray induced ion production in the atmosphere and the corresponding ionization effect were subject of dynamical changes. Here we compute the complex ion production due to cosmic rays during the Halloween sequence of ground level enhancements on October-November 2003 and we estimate the ionization effect. The spectral and angular characteristics of the solar protons are explicitly considered throughout the events as well their time evolution. The ionization effect during the period is computed at several altitudes above the sea level in a region with $R_{c}$ $\le$ 1 GV and $R_{c}$ $\le$ 2 GV.  
\end{abstract}

\maketitle

\thispagestyle{fancy}


\section{INTRODUCTION}
The Earth is continuously  hit by high energy nuclei of galactic origin, known as cosmic rays. They are the main source of ionization
in the troposphere and stratosphere \cite{Bazilevskaya08, Usoskin09}. The contribution of galactic cosmic rays (GCRs) to the atmospheric
ionization is continuous with slight variation in time due to modulation effects in the Heliosphere. Occasionally solar energetic
particles (SEPs) enter the Earth atmosphere, penetrate deep into in the atmosphere or even reach the surface, in a such way leading to ground level enhancements (GLEs). 
As a result they cause an excess of ionization, specifically over the polar caps \cite{Usoskin11a, Mishev2013}. 

At the same time the possible effect of CR induced ionization on various atmospheric processes related to atmospheric chemistry
and physics is debated over the last years. Recent findings suggest an apparent influence of cosmic rays on various atmospheric processes and electric circuit, as well as on minor constituents of the atmosphere \cite{Bazilevskaya08, Mironova2015}. Up to present, in most of the proposed and debated models, the induced by CRs plays a key role. Therefore, study of the induced ionization by GLE particles during some strong events allows one to assess possible effects in enhanced mode.

Nowadays several models based on a full Monte Carlo simulation of the atmospheric cascade are proposed in order to assess the CR induced ionization \cite{Desorgher05, Usoskin06, Velinov2009}. All those models agreed within  10--20 $\%$ \cite{Usoskin09}. These full target models allow one to compute the ion production rate, accordingly ionization effect in the atmosphere during major GLEs as superposition of the contribution of cosmic rays with galactic and solar origin \cite{Mishev13b, Mishev15}, to estimate the ion production rate in a whole atmosphere \cite{Mishev12, Mishev15a} as well as the the corresponding ionization effect \cite{Mishev15b, Mishev16a}. Here we present the results of computation of ion production rate and corresponding ionization effect relative to the average due to GCRs during the Halloween sequence of GLEs on October-November 2003 \cite{Mishev11a}.

\section{MODEL}
Here we use model similar to \cite{Usoskin06}, the full description given elsewhere \cite{Mishev07, Velinov2009}. The ion production rate is given by: 

\begin{equation}
  q(h,\lambda_{m}) =  \frac{1} E_{ion} \sum_{i} \int_{E_{cut}(R_{c})}^{\infty} \int_{\Omega} D_{i}(E) \frac{\partial E(h,E)}{\partial h} \rho(h)dE d\Omega 
        \label{simp_eqn1}
   \end{equation}

\noindent where $\partial E$ is the deposited energy in an atmospheric layer $\partial h$, $h$ is the air overburden above a given altitude in the
atmosphere expressed in $g/cm^{2}$ subsequently converted to altitude above the sea level (a.s.l.), $D_{i}(E)$ is the differential cosmic ray spectrum for a given
component $i$, $\rho$ is the atmospheric density in $g.cm^{-3}$, $\lambda_{m}$ is the geomagnetic latitude, $E$ is the initial energy of the incoming primary nuclei on the top of the atmosphere, $\Omega$ is the geometry factor - a solid angle and $E_{ion}$ = 35 eV is the energy necessary for creation of an ion pair in air \cite{Porter76}. The integration is over the kinetic energy above $E_{cut}(R_{c})$, which is defined by the local rigidity cut-off $R_{c}$ for a nuclei of type $i$ at a given geographic location by the expression:

\begin{equation}
 E_{cut,i}=\sqrt{ \left( \frac {Z_{i}} {A_{i}}\right)^{2} R_{c}^{2}+ E_{0}^{2}} - E_{0}
        \label{simp_eqn2}
   \end{equation}
  
\noindent where $E_{0}$ =  0.938 GeV/n is the proton's rest mass. Accordingly, for SEPs spectra in equation (1), which are considerably varying from event to event, we consider results derived on the basis of ground based measurements with neutron monitors.  In this study, the propagation and interaction of high energy protons with the atmosphere are simulated with the PLANETOCOSMICS code \citep{Desorgher05} assuming a realistic atmospheric model NRLMSISE2000 considering seasonal influence \cite{Picone02, Mishev08, Mishev2014}. PLANETOCOSMICS provides the energy loss and deposition by secondary CRs, necessary for the computations with ~Eq. (\ref{simp_eqn1}). Therefore the model allows one to estimate the ion production rate, accordingly the ionization effect  in a whole atmosphere.

\section{Ion production rate during the Halloween events}
The extreme solar activity in October–November 2003 produced 3 GLEs, with onsets occurring on 28 October, 29 October, and on 2 November. The GLE on 28 October 2003 accompanied a large
flare (4B, X17.2) occurred in the active region NOAA 10486. It occurred during significant interplanetary disturbance related to previously ejected coronal mass ejection (CME)
on 26 October during a 3B/X1.2 flare in the active region AR10486. The Forbush decreases during the sequence of Halloween GLE events was explicitly considered i.e. the GCR flux reduce is taken into account. This is important because the ion production rate during major GLEs is a superposition of the contribution of galactic cosmic rays and GLE particles, which typically  possess an essential anisotropic part. For the computation of ion production rate we assume the force field model of GCR spectrum \cite{Gle68, Cab04} with a solar modulation parameter according to \cite{Usoskin11b}. For the GLE particles we consider compilation of SEP spectra based on neutron monitor reconstructions \cite{Mirosh05}. The ion production rates during the GLE 65 on 28 October 2003 in the polar, sub-polar (rigidity cut-off $R_{c}$ $\le$ 1 GV) and high mid latitudes region (rigidity cut-off $R_{c}$ $\le$ 2 GV) are presented in Fig.1 

 \begin{figure} [!ht]
\centering \resizebox{\columnwidth}{!}{\includegraphics{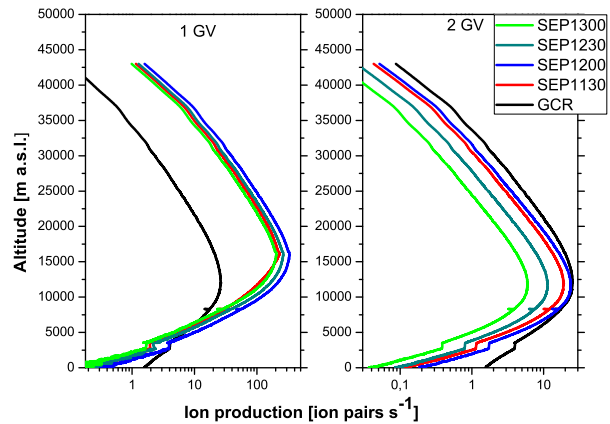}}
 \caption{Ion production rate during the GLE 65 on 28 October 2003 in the polar, sub-polar and high mid latitudes region with rigidity cut-off $R_{c}$ $\le$ 1 GV, accordingly  $R_{c}$ $\le$ 2 GV \label{fig1}}
 \end{figure}

 \begin{figure} [!ht]
\centering \resizebox{\columnwidth}{!}{\includegraphics{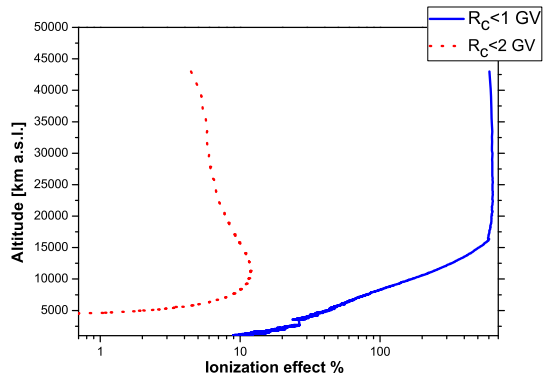}}
 \caption{Maximal ionization effect averaged over the event during GLE 65 on 28 October 2003 in the region with $R_{c}$ $\le$ 1 GV and $R_{c}$ $\le$ 2 GV \ \label{fig2}}
 \end{figure}

The computed ion production rate is significant during the initial and main phase of the event at the polar and sub-polar region with rigidity cut-off of about 1 GV, specifically in the low stratosphere (Fig. 1). The ion production rate remain important during the late phase of the event. In the region of high mid with rigidity cut-off of about 2 GV the ion production is comparable to the average due to GCR. Moreover, at altitudes of about 10 km a.s.l. and below the ion production due to GCR is greater than SEPs, because of the rapidly falling spectra of the solar particles. According our estimation at low mid latitudes with rigidity cut-off of about 3 GV and greater, the ion production due to GCR dominates in the whole atmosphere throughout the event.The corresponding ionization effect relative to ionization due to average of GCR, averaged over the event is shown in Fig.2. The ionization effect is significant (about 200-300 $\%$) and quasi-constant in a whole atmosphere, but low troposphere at the polar and sub-polar region with rigidity cut-off of about 1 GV . In the low stratosphere and troposphere the ionization effect rapidly diminish to about 20 $\%$. In the region of mid latitudes the ionization effect is marginal in a whole atmosphere.

The second Halloween event occurred during a major Forbush decrease and was characterized with smaller increase of NM count rate. Therefore the reduced GCR flux is explicitly considered for the computation. In general this event was weaker than the the GLE 65. In a similar way the ion production rate was computed. The computed ion production rates during the GLE 66 on 29 October 2003 in the polar, sub-polar (rigidity cut-off $R_{c}$ $\le$ 1 GV) and high mid latitudes region (rigidity cut-off $R_{c}$ $\le$ 2 GV) are presented in Fig.3. 

 \begin{figure} [!ht]
\centering \resizebox{\columnwidth}{!}{\includegraphics{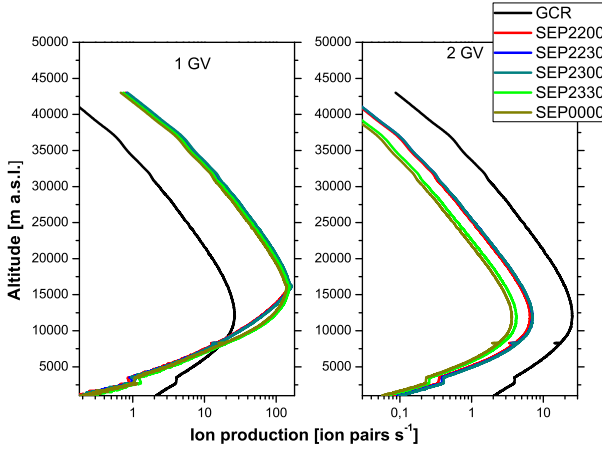}}
 \caption{Ion production rate during the GLE 59 on 28 October 2003 in the polar, sub-polar and high mid latitudes region with rigidity cut-off $R_{c}$ $\le$ 1 GV, accordingly  $R_{c}$ $\le$ 2 GV \label{fig3}}
 \end{figure}

 \begin{figure} [!ht]
\centering \resizebox{\columnwidth}{!}{\includegraphics{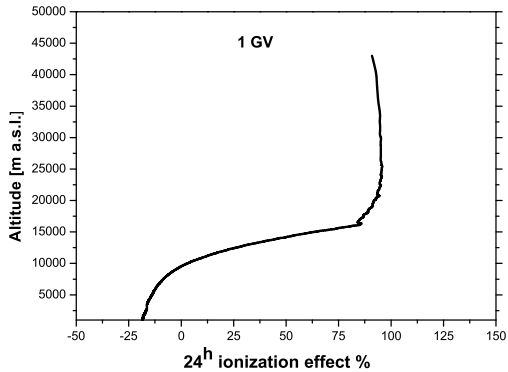}}
 \caption{Maximal ionization effect averaged over the event during GLE 66 on 29 October 2003 in the region with $R_{c}$ $\le$ 1 GV \label{fig4}}
 \end{figure}

The reconstructed from NM data SEP spectra are relatively hard, resulting on significant ion production rate during the event (Fig.3). However, because the strong Forbush decrease and the smaller duration of the event, the corresponding ionization effect was not as significant as in the previous case. The estimated ionization effect is about 90 $\%$. In the troposphere 
it diminish to less than 20 $\%$. The ionization effect is significant only in the polar and sub-polar region with rigidity cut-off of about 1 GV and marginal in the region of mid latitudes. ionization effect is marginal in a whole atmosphere. 
 
\begin{figure} [!ht]
\centering \resizebox{\columnwidth}{!}{\includegraphics{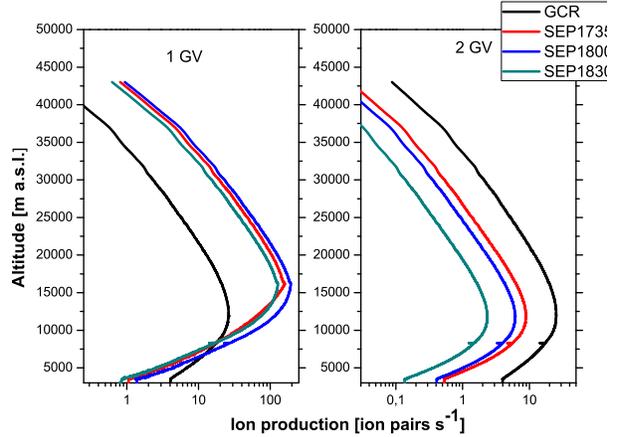}}
 \caption{Ion production rate during the GLE 67 on 2 November 2003 in the polar, sub-polar and high mid latitudes region with rigidity cut-off $R_{c}$ $\le$ 1 GV, accordingly  $R_{c}$ $\le$ 2 GV \label{fig5}}
 \end{figure} 
 
 The GLE 67  event on 2 November 2003 was related to X8.3/2B solar flare. The event onset began at about 17:30 and 17:35 at several
stations and the strongest NM increases was observed at South Pole (36.0 $\%$). The computed ion production rates during the GLE 67 on 2 November 2003 in the polar, sub-polar (rigidity cut-off $R_{c}$ $\le$ 1 GV) and high mid latitudes region (rigidity cut-off $R_{c}$ $\le$ 2 GV) are presented in Fig.5. The ion production rate is significant during the whole event at the polar and sub-polar region with rigidity cut-off of about 1 GV, specifically in the low stratosphere. In the region of high mid latitudes with rigidity cut-off of about 2 GV the ion production is comparable to the average due to GCR.
  
\section{Conclusion}
On the basis of SEP spectra derived from analysis of NM data and with a numerical model based on Monte Carlo simulations we compute the ion production rate and the corresponding effect during the sequence of Halloween GLE events of October–November 2003. It was shown that the ion production rate is significant during the initial and main phase of GLE 65 at the polar and sub-polar region. Accordingly the ionization effect in the polar and sub-polar regions of the Earth  is significant for all events, the strongest during GLE 65. The effect is maximal in the region of the Pfotzer maximum and diminish in the troposphere. In the region of high-middle latitudes, as well as in low- middle latitudes the ionization effect is considerable smaller.

\bigskip 
\begin{acknowledgments}
This work was supported by the CoE ReSoLVE (project No. 272157) of the Academy of Finland.
\end{acknowledgments}

\bigskip 

\end{document}